\documentclass[11pt
]{article} 
\usepackage[utf8]{inputenc}

\usepackage{amsmath}
\usepackage{amssymb}
\usepackage{amsthm} 
\usepackage{ytableau}
\usepackage{multirow}
\usepackage{array}
\usepackage{float}

\usepackage{geometry} 
\usepackage{graphicx} 
\DeclareGraphicsExtensions{.eps,.pdf,.png}
\usepackage{ytableau}
\usepackage{tikz}
\usetikzlibrary{decorations.pathreplacing}
\usepackage{epstopdf}

\usepackage[color,matrix,arrow]{xy}
\xyoption{all}

\usepackage{xstring}
\usepackage{xifthen}

\usepackage{color}

\newcommand{\clb}{\color{blue}}
\newcommand{\cldb}{\color{blue!50!black}}
\newcommand{\clddb}{\color{blue!30!black}}

\renewcommand{\ge}{\geqslant}
\renewcommand{\le}{\leqslant}
\newcommand{\te}{\theta}

\newcommand{\eqn}{\begin{equation}}
\newcommand{\nqe}{\end{equation}}

\newcommand{\vc}[1]{\boldsymbol{\mathsf #1}}

\newcommand{\vx}{\boldsymbol{x}}

\newcommand{\NN}{\mathbb{N}}

\newcommand{\GG}{\Gamma}
\newcommand{\CC}{\mathcal{C}}

\newcommand{\Dir}{\mathcal{D}}

\newcommand{\tpPD}{two-parameter Poisson-Dirichlet } 
\newcommand{\prg}[1]{\paragraph{\cldb #1}}

\newcommand{\al}{\alpha}

\newcommand*\dif{\mathop{}\!\mathrm{d}}

\newcommand{\vcc}[2]{\vc {#1}=({#1}_1,\dots,{#1}_{#2})}

\newcommand{\summ}[3]{\sum_{#1=#2}^#3}

\newcommand{\fm}[2]
{
  \StrLen{#2}[\MyStrLen]
  \ifthenelse{\equal{\MyStrLen}{1}}
	{\frac {#1^{#2}}{#2!}}
	{\frac {{#1}^{{#2}}}{({#2})!}}
}

\newcommand{\lif}[2]
{
  \StrLen{#2}[\MyStrLen]
  \ifthenelse{\equal{\MyStrLen}{1}}
	{\frac {#1_#2}{#2!}}
	{\frac {#1_{#2}}{(#2)!}}
}

\newcommand{\rfm}[2]
{
  \StrLen{#1}[\argum]
  \StrLen{#2}[\expon]
	\ifthenelse{\equal{\argum}{1}}
		{\ifthenelse{\equal{\expon}{1}}
			{\frac {#1^{[#2]}}{#2!}}
			{\frac {#1^{[#2]}}{(#2)!}}}
		{\ifthenelse{\equal{\expon}{1}}
			{\frac {(#1)^{[#2]}}{#2!}}
			{\frac {(#1)^{[#2]}}{(#2)!}}}
}

\newcommand{\ffm}[2]
{
  \StrLen{#1}[\argum]
  \StrLen{#2}[\expon]
	\ifthenelse{\equal{\argum}{1}}
		{\ifthenelse{\equal{\expon}{1}}
			{\frac {#1_{(#2)}}{#2!}}
			{\frac {#1_{(#2)}}{(#2)!}}}
		{\ifthenelse{\equal{\expon}{1}}
			{\frac {(#1)_{(#2)}}{#2!}}
			{\frac {(#1)_{(#2)}}{(#2)!}}}
}

\newcommand{\ffl}[2]
{
  \StrLen{#1}[\MyStrLen]
  \ifthenelse{\equal{\MyStrLen}{1}}
	{{#1}_{({#2})}}
	{({#1})_{({#2})}}
}

\newcommand{\rfl}[2]
{
  \StrLen{#1}[\MyStrLen]
  \ifthenelse{\equal{\MyStrLen}{1}}
	{{#1}^{[{#2}]}}
	{({#1})^{[{#2}]}}
}

\newcommand \exty {exchangeability }

\newcommand \dn {{\sc dn}}
\newcommand \up {{\sc up}}

\newcommand \vsk {\vskip 1mm}

\renewcommand \prg[1] {\vskip 0.05cm  {\cldb \bf {#1}}}

\usepackage{capt-of}
\usepackage{mathtools} 
\usepackage{subcaption}
\usepackage{booktabs}

\usepackage{vmargin} 
\setpapersize{A4}
\setmargins
		{1.1cm}
		{0.7cm}
         	{19.0cm}
         	{24.5cm}
         	{12pt}
         	{1cm}
         	{0pt}
         	{1.25cm}

\title{Market shape formation, statistical equilibrium \\
and neutral evolution theory 
}
\author{  Sergey Sosnovskiy
 \\\tt \small ssnv.sky@gmail.com
}
\small \date{\today} 

\begin{document}
\maketitle
\abstract{ 
Mathematical methods of population genetics and framework of exchangeability provide a Markov chain model for analysis and interpretation of stochastic behaviour of equity markets, explaining, in particular, market shape formation, statistical equilibrium and temporal stability of market weights.
}

\section{Introduction}
\vskip -2mm
Log-log plot of normalized stock market capitalizations ranked in descending order is called {\em capital distribution curve}. 
For example, figures below display distribution of capital on the NASDAQ
market on three dates in 2014 (data source is {\tt http://www.google.com/finance\#stockscreener}). 
Ranked market weights experienced relatively small fluctuations, despite significant changes in overall capitalization of the NASDAQ market during that period of time. 

\vskip 0.1cm
\begin{figure}[H]
\centering
\includegraphics[width=0.65\linewidth, height=5.5cm,
trim=2.1cm 10.3cm 2cm 11.75cm,clip]{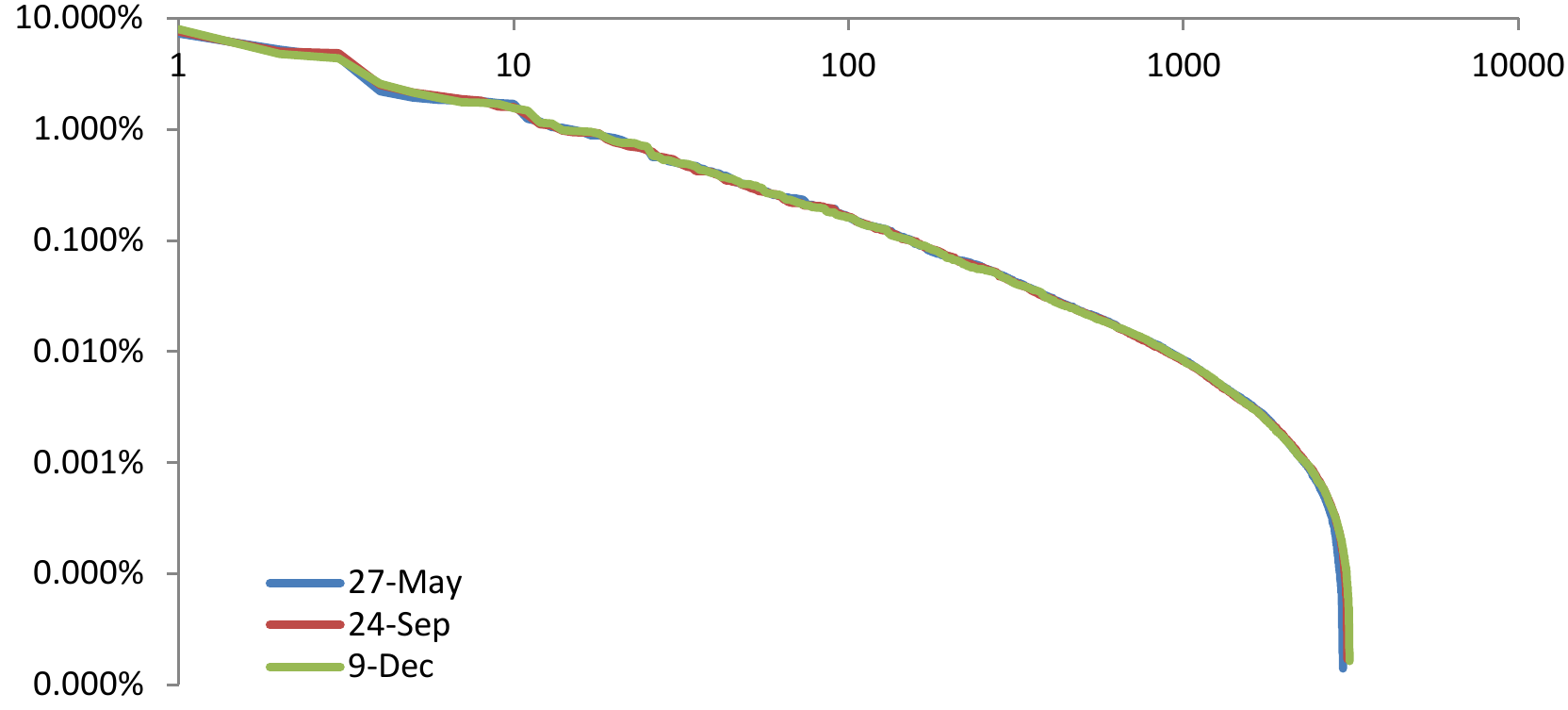}	

\end{figure}
\vskip -0.15cm
\begin{figure}[H]
\centering
\includegraphics[width=0.65\linewidth, height=5.5cm,
trim=2.1cm 10.25cm 2cm 11.75cm,clip]{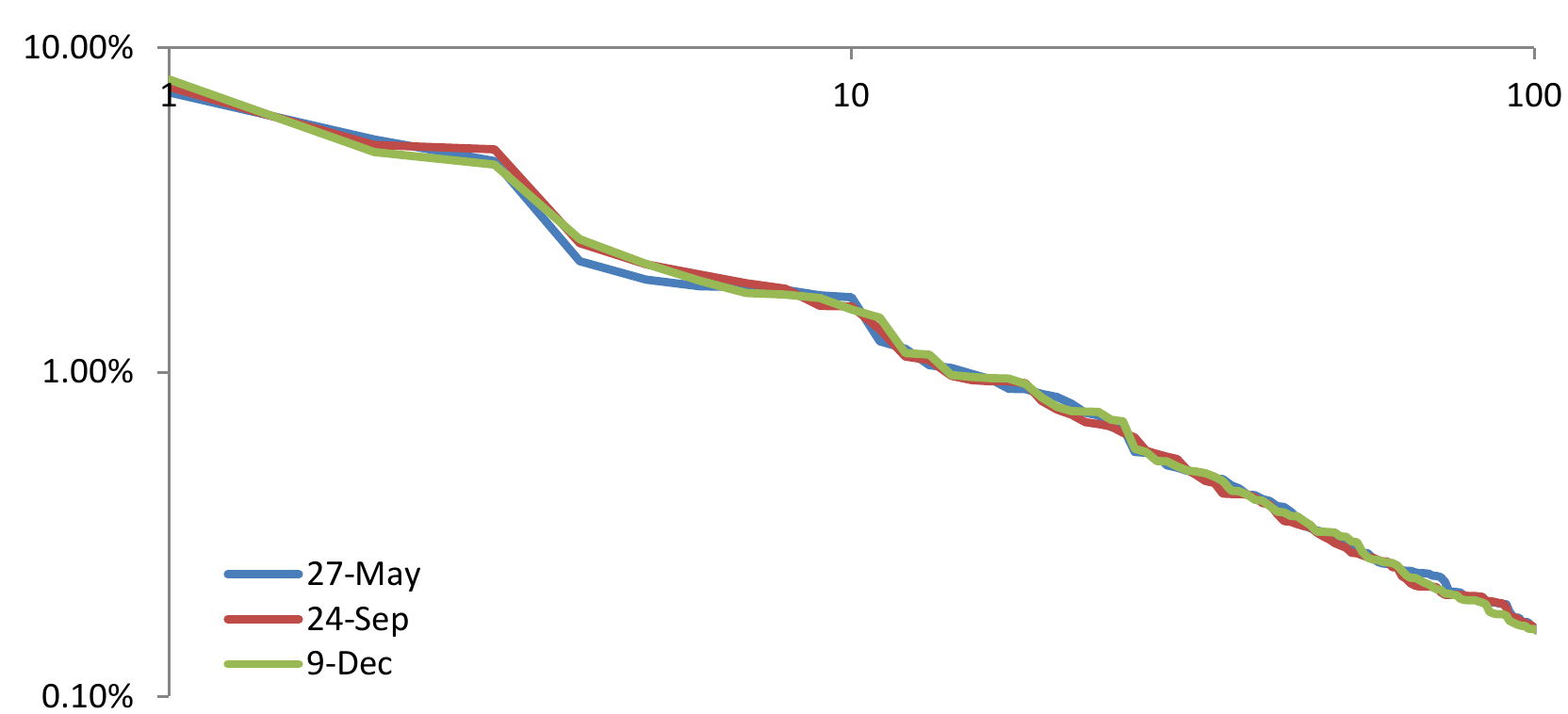}
\caption{\small NASDAQ capital distribution curves, all stocks  (above) and top 100 stocks (below)}\label{fig3}
\end{figure}
One of the aims of this paper is to provide an example of a possible mechanism explaining
{\bf\em temporal stability} and 
{\bf\em statistical equilibrium} 
of 
normalized stock capitalizations 
by means of the Polya-Dirichlet Markov chain, 
analogous to the Wright-Fisher model 
of neutral theory of evolution.
\vsk
\prg{Classic and neutral evolution theory.}
Classic form of Darwinian theory suggests that forces of natural selection play central role in evolution of species. Theory of neutral evolution, proposed by Kimura, complements the classic theory by adding genetic dimension. 
Kimura observed that discrepancies in traits, 
such as 
 small variations in colouring of beaks or feathers in a population of birds, 
occur at molecular-genetic level due to random effects in reproduction and majority of these variations are neutral with respect to fitness. 
According to the neutral theory, force of natural selection is still be important since it purges deleterious mutations. However, majority of surviving mutations are neutral, and possibly only few are advantageous. 


Mutations and random combinations of genes in new generations
lead to fluctuations of allelic frequencies or {\em genetic drift}. 
The Wright-Fisher and Moran models describe stochastic evolution of genetic frequencies 
as statistical equilibrium fluctuations, modelled by 
diffusion process with stationary Dirichlet distribution.


\vsk
\prg{Evolution theory and finance.} Application of evolutionary ideas in finance has a long history dating back to Malthus, Marshall and many others. 
Recently Evstigneev, Hens, and Schenk-Hopp{\'e} \cite{evstigneev2008evolutionary} developed descriptive model of Evolutionary Finance, which employs principle of natural selection for modeling dynamics of asset prices and
analysis of investment strategies.


Kirman 
\cite{kirman1993ants} 
considered 
version of the Wright-Fisher model with mutation in a context of economic interpretation of behavior of ants searching for a food source. He
observed that proportion of ants choosing one of the possible food channels is better described by stationary distribution of a Markov chain, rather than by single point of equilibrium. He proposed that 
the 'herding' behaviour on financial markets 
as well is better described by means of stochastic equilibrium, rather than by single or multiple equilibria.


\vsk
\prg{Formation of market limit shape.} Standard and non-linear versions of the Polya process  have been used by Arthur et al. \cite{arthur1994increasing} for illustration of 
appearance of market structure. 
Polya scheme has the following interesting property: proportions of balls converge to some limiting values, but these limits are random and described by the Dirichlet distribution.



\vsk
\prg{Markov lattice and reversibility.} 
Polya-Dirichlet Markov chain with state space defined on  
lattice of ordered integer partitions provides a framework for analysis and modeling of  
stochastic equilibrium of market weights. 
Transitions on the lattice of partitions are described in terms of 
random \up- and \dn- operators proposed by Kerov \cite{kerov2003asymptotic},  Fulman\cite{fulman2005stein}, Borodin and Olshanski \cite{borodin2009infinite} and Petrov \cite{petrov2007two}. 
Historically, Markov chains with \dn/\up-transitions in a context of Polya model were first studied by Costantini, Garibaldi, et al. in \cite{costantini2000purely}, \cite{garibaldi2004finitary}. 

\vsk
\prg{Exchangeability and random fluctuations.}
Infinite \exty implies existence of \up- and \dn- random transitions, connecting adjacent levels of integer compositions. It is shown in Section \ref{stocheq} that probabilities of these transitions satisfy reversibility conditions and therefore induce a lattice of Markov chains. 
Random transitions on this lattice correspond to statistical equilibrium behaviour of market weights or allelic frequencies not only for fixed, but also for varying market or population sizes.

\vsk
\prg{Neutral theory and financial markets.}
The Polya-Dirichlet Markov lattice corresponds to the discrete version of the Wright-Fisher process with mutations and provides a toy model of equilibrium markets behaviour.

\begin{itemize}
	\item After initial phase of rapid expansion, 
in a same way as proportions of balls converge to random limits in Polya scheme, 
market weights settle down and form capital distribution curve. 
	\item Up- and down- changes in overall market capitalization lead to random drift of market weights fluctuating in stochastic equilibrium around limiting values, given by the capital distribution curve. The stationary distribution of market weights can be modeled by means of the \up- and \dn- Markov chain.
	\item In general, increase of market capitalization enforces market structure and decrease of capitalization leads to weakening of the structure and higher volatility, which creates an opportunity for market reshaping. This mechanism is analogous to the so-called {\em nearly neutral theory of evolution} proposed by Ohta \cite{ohta1992nearly}, in which smaller populations have faster molecular-genetic evolution and adaptation rate.
	\item {\clddb This theory provides interpretation of market crises as markets self-adaption to changing economic conditions, where capitalization decrease leads to market reshaping and faster adjustment to new econo-financial landscape.
	\item Arbitrage opportunities can be considered as corresponding to deleterious mutations, eliminated by forces of natural selection.}
\end{itemize}

\newpage
\prg{Mechanics, economics and reversible equilibrium.} 
As pointed out by Garibaldi and Scalas \cite{garibaldiscalas2010}, equilibrium modeling in economics and finance was developed under strong influence of ideas of static or classical mechanics. 
Alternative approach  is provided by framework of stochastic equilibrium and reversibility  conditions, which have roots in Boltzmann's work on statistical mechanics. Exhaustive treatment of econophysics from the point of view of \exty is contained in the book of Garibaldi and Scalas \cite{garibaldi2010finitary}.

Excellent explanation of the framework of reversible equilibrium 
is contained in the classic book of Kelly \cite{kelly2011reversibility}.


\section{Polya process with down/up transitions}
In a classic form of Polya process colored balls are placed into a box with probabilities proportional to weights of balls of existing colors. The process provides a discrete counterpart of the Dirichlet distribution, since if vector $(\al_1,...,\al_m)$ represents  initial/prior weights of balls of each color, limiting values of proportions of weights in Polya scheme have Dirichlet distribution with the same vector of parameters $\Dir_m (\al_1,...,\al_m)$. 

Modified Polya process, in which balls can also be removed illustrates  important ideas of
\begin{itemize}
	\item appearance and temporal stability of ranked proportions, and 
	\item stochastic equilibrium of these weights.
\end{itemize} 
Let us consider an artificial stock market with finite number of stocks represented by $m$ different colours. Initially in the box there are $m$ 'prior' balls of each colour and the same weight $\al$, such that total weight of all balls is $\te=m\al$. In other words, all stocks start with the same initial conditions and colours (or tickers) are used only to distinguish the stocks. 
Vector $\vcc n m$ represents {\em stock capitalizations} equal to number of {\em placed} balls of each color at stage $n=n_1+...+n_m$, so at initial stage this vector is $\vc n=(0,...,0)$. 
All possible market configurations with overall capitalization $n$ are represented by compositions (ordered partitions) in the integer-valued simplex
\[
	\CC_n=\big\{\vc n=(n_1,...,n_m) \;\big|\; n_i \in\NN_0,\; 
	\textstyle\sum n_i=n \big\}
\]

At the first step one of the prior balls is drawn with probability $\al/\te=1/m$. 
The ball is placed back into the box together with a ball of the same color and unit weight.  At stage $n$ probability to add a ball of color $i$
 is 
\[
p=\frac{\al+n_i}{\te+n},
\]
where $n_i$ denotes number of balls of color $i$ in the box. 
For instance, with $m=3$ colors, say red, green and blue, probability of drawing 3 red balls, 2 green ones and 1 blue ball in this particular sequence is
\[
	p('rrrggb')=\frac{\al(\al+1)(\al+2)}{\te(\te+1)(\te+2)}
	\cdot \frac{\al(\al+1)}{(\te+3)(\te+4)}
	\cdot \frac{\al}{\te+5}
	=\frac{\al^{[3]}\al^{[2]}\al^{[1]}}{\te^{[6]}},
\]
with raising or ascending factorial power defined as\vskip -0.2cm
\[\rfl \al k=\al(\al+1)\cdots(\al+k-1)=\frac{\GG(\al+k)}{\GG(\al)}\] 
By combinatorial argument probability of configuration $\vc n=(n_1,...,n_m)$ at stage $n$ is 
\eqn\label{Ply}
	p(\vc n)=\binom n{n_1,...,n_m} \frac{\prod_{i=1}^m \rfl \al{n_i}}{\rfl \te n}
	=\frac{n!}{\rfl \te n} \frac{\al^{[n_1]}}{n_1!}\cdots\frac{\al^{[n_m]}}{n_m!}
\nqe
For each level $n$ this formula establishes probability distribution in the simplex $\CC_n$, moreover this distribution is {\em exchangeable} or symmetric, in a sense that probability of any sequence does not depend on the order of balls drawn and depends only on the number of balls of each color.

Using asymptotic $\fm \al n \asymp \frac {n^{\al-1}}{\GG(\al)}$ 
\[
	p(\vc n)\asymp \frac{\GG(\te)}{n^{\te-1}}
	\frac {n_1^{\al-1}}{\GG(\al)} \dots \frac {n_m^{\al-1}}{\GG(\al)}
	=\frac{\GG(\te)}{(\GG(\al))^m} \prod_{i=1}^m \Big(\frac{n_i}n\Big)^{\al-1} 
	\cdot  \Big(\frac1n\Big)^{m-1}
\]
which corresponds to density of symmetric Dirichlet distribution
$
f_{\vc \al}(\vx)\dif \vx=\frac{\GG(\te)}{(\GG(\al))^m}\; x_1^{\al-1} \cdots x_m^{\al-1} \dif x_1 \cdots \dif x_{m-1}$

\newpage
\prg{\up- and \dn- transitions.} 
In a standard Polya model number of balls increases
at each stage, such that in configuration $\vcc nm \in \CC_n$ component $i$ increases by one with conditional probability 
\eqn\label{pr-up}
	{\mathsf u}_{i,n}=\frac{\al+n_i}{\te+n}
\nqe
This can be considered as random {\up}-transitions of configuration from simplex $\CC_n$ to $\CC_{n+1}$. In financial terms, {\up}-moves correspond to investment into particular stock and increase of capitalization. Clearly, stochastic dynamics of these transitions is of preferential attachment type, since conditional probability \eqref{pr-up} of stock growth is proportional to its capitalization $n_i$. As shown below \up-moves preserve probability distributions \eqref{Ply} on simplexes $\CC_n$.  

It turns out, that \up-moves also implicitly define \dn-transitions, which randomly move configuration backwards from simplex $\CC_n$ to $\CC_{n-1}$. These \dn-transitions also preserve probability structure on simplexes. In terms of Polya's model \dn-move corresponds to
removing a ball of some color at random and financially it has interpretation of decrease of capitalization of one of the stocks by one unit.

Structure of exchangeable probability distribution plays an important role connecting \up- and \dn- transitions, dual to each other. For the sake of illustration let us consider case of two stocks, labelled by two colours. Structure of connections of probabilities between simplexes $\CC_0,\CC_1,\CC_2,...$ with $m=2$ is shown below, where $p_{k,n-k}$ denotes probability of configuration with $k$ balls of the first color and $n-k$ balls of the second color. 
\[\xymatrixcolsep{3pc}\xymatrixrowsep{0.01pc}
\xymatrix{  
				 		&					  					& p_{2,0} \\
				 		& p_{1,0}\ar@{-}[ur]\ar@{-}[dr] 	&  			 \\
p_{0,0}\ar@{-}[ur]\ar@{-}[dr]&							& p_{1,1} \\
						& p_{0,1}\ar@{-}[ur]\ar@{-}[dr] 	& 			 \\
						&										& p_{0,2}
}
\]
Let us consider probability flows between levels $\CC_{n-1}$ and $\CC_n$ for $n_1+n_2=n$
\vskip -0.2cm
\[\xymatrixcolsep{4pc} \xymatrixrowsep{0.45pc}
\xymatrix{  
													& {p_{n_1+1,n_2-1}} \\
 p_{n_1,n_2-1}\ar[ur]\ar@[blue][dr]	|{\clb \frac{\al+n_2-1}{\te+n-1}}&\\
													& {\clb p_{n_1,n_2}} \\
 p_{n_1-1,n_2}\ar[dr]\ar@[blue][ur]|{\clb \frac{\al+n_1-1}{\te+n-1}} &\\
 													& p_{n_1-1,n_2+1}
}
\]
For instance, configuration $(n_1,n_2-1)$ 
can migrate to state $(n_1+1,n_2-1)$ with probability $\frac{\al+n_1}{\te+n-1}$
or go to state $(n_1,n_2)$ with probability $\frac{\al+n_2-1}{\te+n-1}$, thus   contribution or {\em forward probability flow} to configuration $(n_1,n_2)$ is
\[
	\mathsf{pf}_{(n_1,n_2-1) \to (n_1,n_2)}=p_{n_1,n_2-1}\frac{\al+n_2-1}{\te+n-1}
\]
Similarly it can be shown that probability flow from state $(n_1-1,n_2)$ to state $(n_1,n_2)$ is
\[
	\mathsf{pf}_{(n_1-1,n_2) \to (n_1,n_2)}=p_{n_1-1,n_2}\frac{\al+n_1-1}{\te+n-1}
\]
It is easy to see that \up-moves preserve probability measure \eqref{Ply} on simplexes. For instance, total flow of probabilities into state $(n_1,n_2)$ is
\[\textstyle
	\binom{n-1}{n_1}\frac{\al^{[n_1]}\rfl \al{n_2-1}}{\rfl \te {n-1}}
	\cdot \frac{\al+n_2-1}{\te+n-1}+
	\binom{n-1}{n_1-1}\frac{\al^{[n_1-1]}\rfl \al{n_2}}{\rfl \te {n-1}}
	\cdot \frac{\al+n_1-1}{\te+n-1}
	=\big(\binom{n-1}{n_1}+\binom{n-1}{n_1-1}\big) 
		\frac{\rfl \al {n_1} \rfl \al{n_2}}{\rfl \te {n}}
	=\binom n{n_1,n_2} \frac{\rfl \al {n_1} \rfl \al{n_2}}{\rfl \te {n}}
\]
In other words, fraction 
$
	\binom{n-1}{n_1,n_2-1}\Big/\binom{n}{n_1,n_2}=\frac {n_2}n
$
of probability $p_{n_1,n_2}$ comes from configuration $(n_1,n_2-1)$ and fraction $\binom{n-1}{n_1-1,n_2}/\binom{n}{n_1,n_2}=\frac {n_1}n$ comes from configuration $(n_1-1,n_2)$. This means that given probability of state $p_{n_1,n_2}$ {\em backward probability flow} can be interpreted as random removal of one of the balls with $p=n_i/n$: 
\vskip -0.2cm
\[\xymatrixcolsep{4pc} \xymatrixrowsep{0.45pc}
\xymatrix{  
{}_{(n_1,n_2-1)}&\\
& {\clb {}_{(n_1,n_2)}}\ar@{.>}[ul]|{\frac{n_2}n}\ar@{.>}[dl]|{\frac{n_1}n} \\
{}_{(n_1-1,n_2)}\\
& 
}
\]
In general, random {\sc down}-transition moves configuration $\vc n=(n_1,...,n_m)$ from simplex $\CC_n$ to $\CC_{n-1}$, and in terms of Polya model it removes a ball of color $i$ with probability 
\eqn\label{pr-dn}
\mathsf d_{i,n}=\frac{n_i}n
\nqe
It is straightforward to show that \dn-moves also preserve probability measure \eqref{Ply}.
In terms of artificial market model \up- and \dn- transitions correspond to increase and decrease of capitalization  (buying/selling) of a stock by one unit of money. 

\section{Market shape formation}
Classic Polya scheme (with \up-transitions) provides a model for simulation of market growth. It turns out that in this model, even if initial conditions are the same for all stocks, over certain period of time, or after reaching certain market capitalization, power-law shaped market structure begins to appear.

Figures below illustrate shape formation of the capital distribution curve on the artificial market. Left side of figures contains realization of market weights, modeled by \up-transitions and right side displays log-log plot of ranked weights at the terminal stage.
All stocks have the same initial conditions modeled by equal prior $\al$. 
After rapid period of 'Big Bang'-like chaotic expansion, 
market weights 
begin to settle down and after step 1500 do not change significantly. 

\iftrue

\begin{figure}[H]\label{f-3}
\centerline{
\includegraphics[width=17.cm, height=6.0cm,
trim=0.6cm 8.5cm 0.6cm 9cm,clip]
{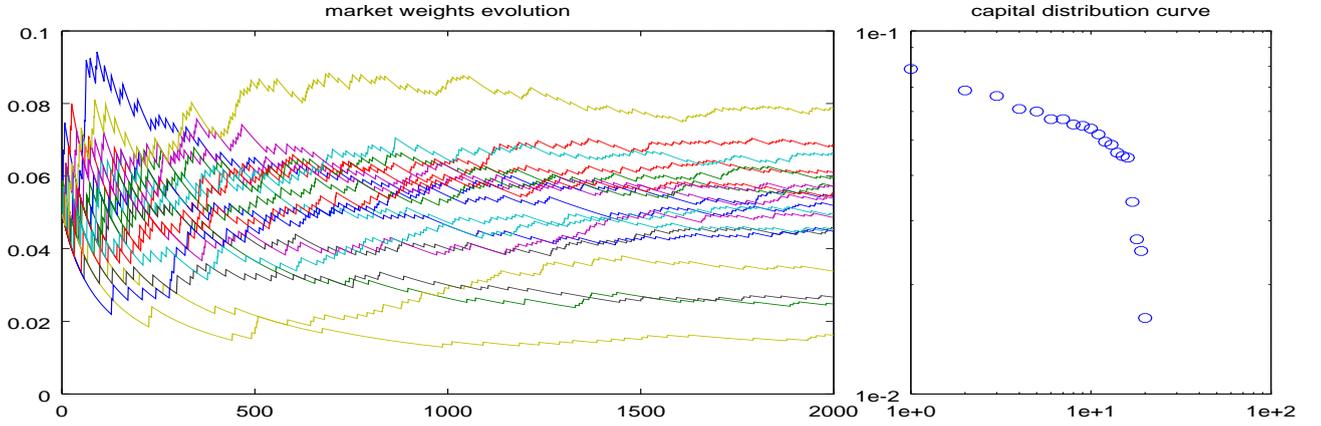}}
\caption{\small Dynamics of market weights and capital distribution, 20 stocks  $(\al=5,\te=100)$}
\end{figure}
Next figure illustrates that for smaller values of parameter $\al$ there is greater variation of market weights. Particular choice of parameter $\al=1$ corresponds to the uniform stick-breaking niche model. 
\begin{figure}[H]\label{f-4}
\centerline{
\includegraphics[width=17.cm, height=6.0cm,
trim=0.6cm 8.5cm 0.6cm 9cm,clip]
{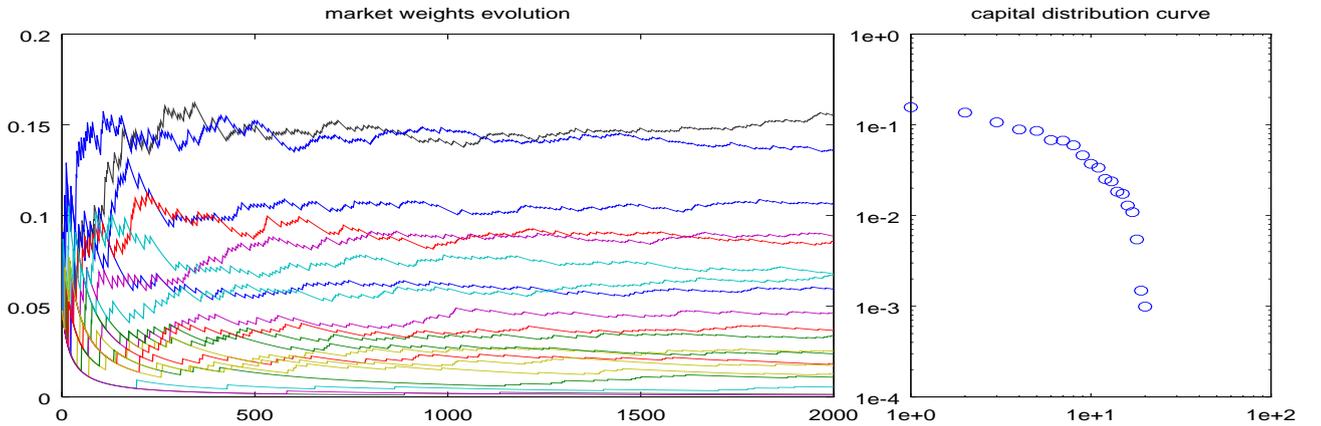}}
\caption{\small Dynamics of market weights and capital distribution,  20 stocks $(\al=1,\te=20)$}
\end{figure}
\fi

\newpage
\section{Statistical equilibrium and Polya-Dirichlet Markov process}\label{stocheq}
One of the central ideas of neutral evolution theory is that proportions of genes in a population (allelic frequencies) experience {\em random drift}, in other words they fluctuate around some values. It is important that population size may increase or decrease (in reasonable amounts), but allelic frequencies remain approximately the same.
The same phenomenon is observed on financial markets: ranked equity capitalizations, comprising capital distribution curve display remarkable temporal stability, despite significant fluctuations of overall capitalization. 

Wright-Fisher model (WF) and its generalizations provide a framework for modeling evolution of proportions of genes fluctuating in stochastic equilibrium. 
Finite form of the $m$-allele WF-model is based on a Markov chain with state space of ordered integer partitions (compositions) with $m$ elements. 
Since stationary distribution of proportions in limiting case of the WF-model with mutation is given by Dirichlet distribution, for consistency 
it is assumed that finite version of WF-model is approximated by the Polya distribution \eqref{Ply}.


In Polya model with \dn/\up-transitions ordered partition 
$\vcc nm$ 
may represent: 
\begin{itemize}
\item vector of stock capitalizations with overall market capitalization $n$,
\item number of genes of each specific type in a population of size $n$.
\end{itemize}
As above it is assumed that vector of priors (or mutation rates, correspondingly), is a symmetric vector with all components equal to $\al$ and $\te=m\al$ denotes sum of all parameters. 
Stationary distribution with transitions in integer simplex $\CC_n$ can be constructed by combining \dn- and \up-transitions, as proposed in \cite{borodin2009infinite},\cite{fulman2005stein},\cite{petrov2007two} and \cite{garibaldi2004finitary}. 
For instance, transition where one item moves from category $i$ to category $j$ in \dn/\up-scheme, is modeled in two steps. 
\[
(..,n_i,..,n_j,..) 
\longmapsto (..,n_i-1,..,n_j,..) 
\longmapsto (..,n_i-1,..,n_j+1,..) 
\]
with probability $q_{i \mapsto j}=\frac{n_i}{n}\frac{n_j+\al}{n+\te-1}$ for $i\neq j$, and with $q_{i \mapsto i}=\frac{n_i}{n}\frac{n_i+\al-1}{n+\te-1}$ as probability
of return. 
\vsk
\prg{Reversibility and stochastic equilibrium.}

If $q(\vc a \mapsto \vc b)$ denotes conditional probability of migration from state $\vc a$ to state $\vc b$, then 
reversibility (detailed balance) conditions imply for all states 
in state space $\mathcal S$ 
\eqn\label{d-balance}
	p(\vc a)q(\vc a \mapsto \vc b)=p(\vc b)q(\vc b \mapsto \vc a), 
	\qquad \forall \vc a, \vc b \in \mathcal S
\nqe
Besides time-reversibility, there are some other interesting interpretations of detailed balance conditions.
\begin{itemize}
	\item In mechanical systems if each process is matched by its reverse process, then the system is in equilibrium. Equilibrium may take place on a micro-level, while on a macro-level the system may stand still.
	\item In terms of probability flows, essentially \eqref{d-balance} states that unconditional probability flow from $\vc a$ to $\vc b$ must be equal to probability flow from state $\vc b$ to state $\vc a$. 
\end{itemize}

It is easy to show that both \dn/\up- and \up/\dn- schemes satisfy detailed balance conditions and therefore they induce reversible and hence stationary Markov chain with a state space of ordered partitions $\CC_n$ for each $n$.

More detailed analysis reveals that reversibility conditions connect probability distributions on all adjacent simplexes $\CC_{n-1}$ and $\CC_n$
\vskip -0.4cm
\[\xymatrixcolsep{4pc} \xymatrixrowsep{0.35pc}
\xymatrix{  
					& {(n_1+1,n_2-1)}\ar@{.>}[dl]|{} \\
 (n_1,n_2-1)\ar@<0.75ex>[ur]|{} \ar@<0.75ex>[dr]|{}	&\\
					& {\;\;(\;n_1\;,\;n_2\;)}\ar@{.>}[dl]|{}\ar@{.>}[ul]|{} \\
(n_1-1,n_2)\ar@<0.75ex>[dr]|{}\ar@<0.75ex>[ur]|{} &\\
 					& (n_1-1,n_2+1)\ar@{.>}[ul]|{}
}
\]
For instance, probability flows between compositions $(n_1-1, n_2)$ and $(n_1, n_2)$ satisfy
\[
	\frac{(n-1)!}{\rfl \te {n-1}} \frac{\rfl \al{n_1-1}\rfl \al{n_2}}{(n_1-1)!n_2!} 
	\cdot {\clb \frac{n_1-1+\al}{n+\te-1}}=
	\frac{n!}{\rfl \te n} \frac{\rfl \al{n_1}\rfl \al{n_2}}{n_1!n_2!} 
	\cdot {\clb \frac{n_1}{n}},
\]
which also clarifies role of \up- and \dn- operators.
In other words, probability distributions \eqref{Ply} on simplexes are pairwise connected. Since for sufficiently large $n$ Polya distribution approximates Dirichlet distribution, sequences of \up- and \dn-operators between simplexes approximate equilibrium process with stationary Dirichlet distribution even with changing values of $n$.

This provides a framework for modeling evolution of market weights for fixed and variables values of $n$. 
It turns out that behavior of equilibrium process depends on the population size. 
When overall market capitalization is sufficiently large 
market structure becomes enforced. 
In contrast, decrease of market cap leads to higher variations of market weights.

For instance, figures below illustrate two phases of evolution of market weights. During the first phase, for $t\le500$ (Figure 5) or $t\le1500$ (Figure 6), the market experiences period of growth and formation of limiting weights, representing capital distribution curve,  displayed on the right subplots with blue circles. Once market value reaches threshold value of $n=500$ or $n=1500$ market capitalization begin fluctuating in \dn/\up-transitions, which generates stochastic evolution of market weights. Corresponding capital distribution curve at the terminal period is shown with red circles. As it can be seen from the figures, stocks with larger capitalizations have higher trading activity.

\begin{figure}[H]
\centerline{
\includegraphics[width=17.25cm, height=6.cm]{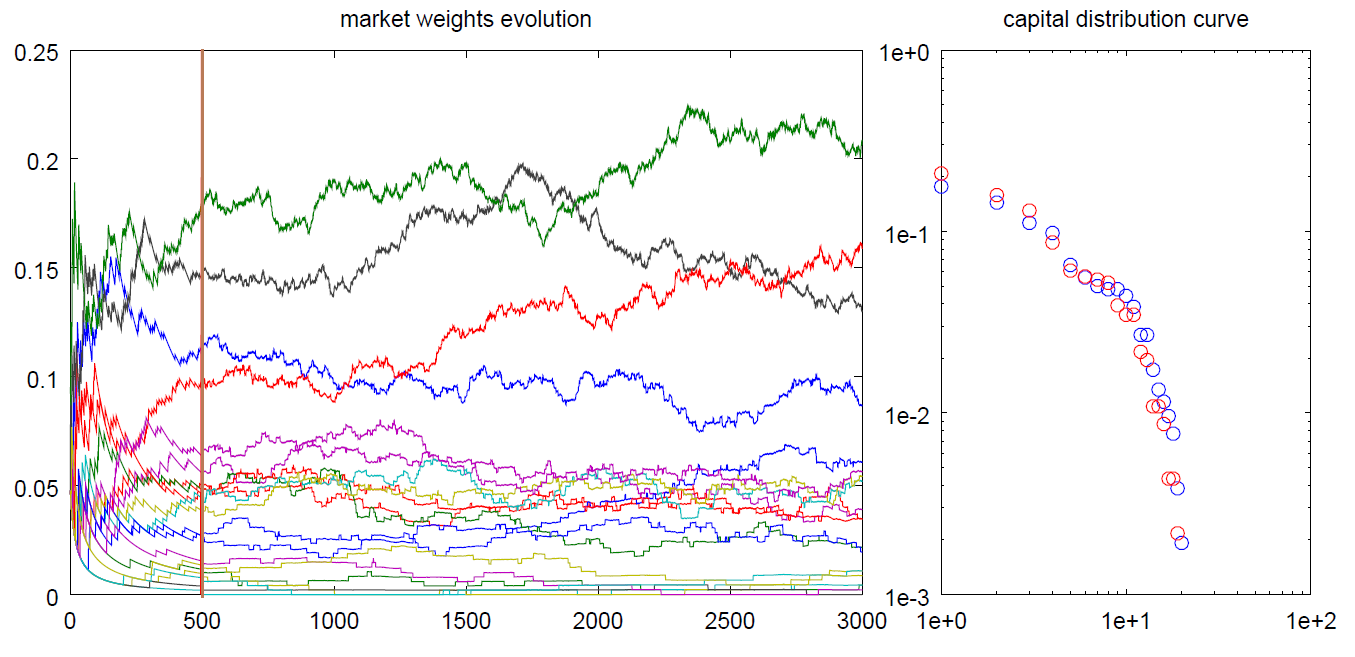}}
\caption{\small Fluctuations at level $n=500$ begin at $t=500$, $(\al=1,\te=20)$, 20 stocks}
\end{figure}

\begin{figure}[H]
\centerline{
\includegraphics[width=17.25cm, height=6.cm]{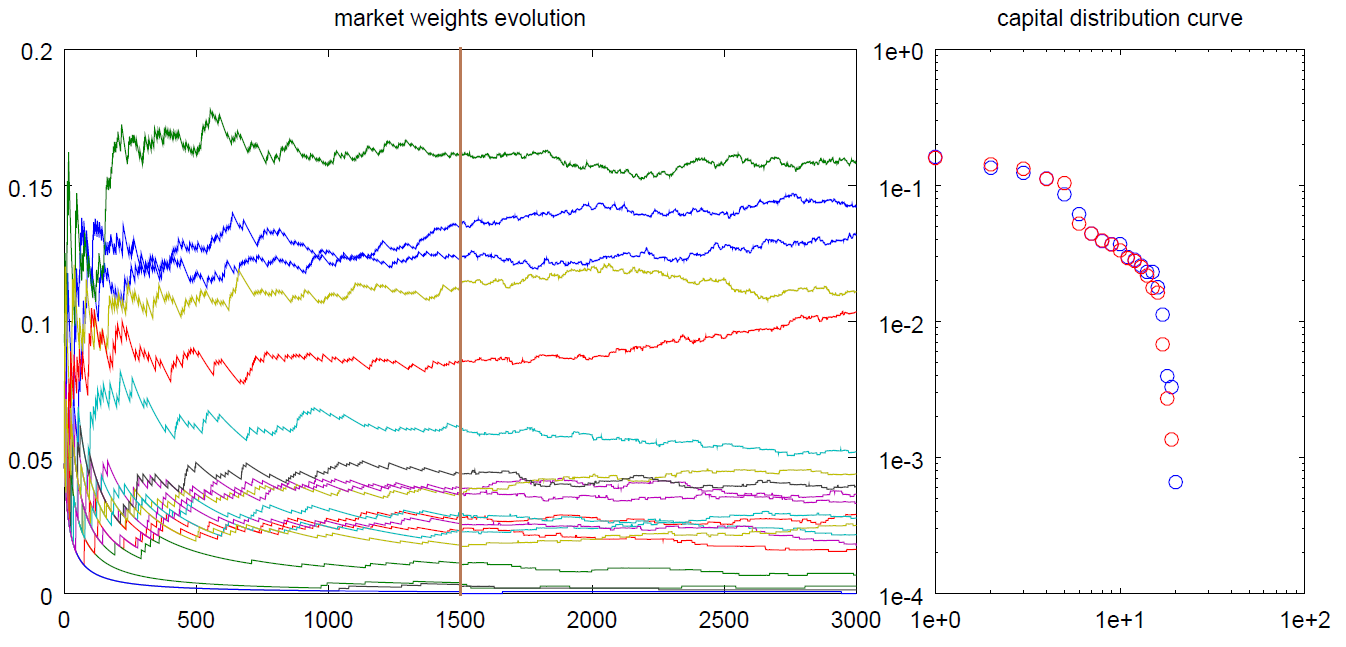}}
\caption{\small Fluctuations at level $n=1500$ begin at $t=1500$, $(\al=1, \te=20)$,  20 stocks
}
\end{figure}

Such behavior of weights, dependent on level $n$ is consistent with the so-called 'nearly neutral theory', developed by Ohta \cite{ohta1992nearly}, where it is proposed that smaller populations experience faster rate of genetic evolution. 
In financial terms, it suggests the market crises can be 'explained' as markets self-adaption to changing economic conditions, where reduced market capitalization allows faster reshaping and adjustment of capitalization structure to new econo-financial landscape.


\newpage
\bibliographystyle{abbrv}
\bibliography{REFF}

\end{document}